\def\ptitle{Spectral bounds for the Hellmann potential}
\input psfig.sty 
\nopagenumbers
\magnification=\magstep1
\hsize 6.0 true in 
\hoffset 0.25 true in 
\emergencystretch=0.6 in                 
\vfuzz 0.4 in                            
\hfuzz  0.4 in                           
\vglue 0.1true in
\mathsurround=2pt                        
\topskip=24pt                            
\def\nl{\noindent}                       
\def\np{\hfil\vfil\break}                
\def\ppl#1{{\leftskip=9cm\noindent #1\smallskip}}    
\def\title#1{\bigskip\noindent\bf #1 ~ \tr\smallskip} 
\font\tr=cmr10                          
\font\bf=cmbx10                         
\font\sl=cmsl10                         
\font\it=cmti10                         
\font\trbig=cmbx10 scaled 1500          
\font\tiny=cmr8                         
\def\ng{>\kern -9pt|\kern 9pt}          
\def\hi#1#2{$#1$\kern -2pt-#2}          
\def\hy#1#2{#1-\kern -2pt$#2$}          

\def\half{{1 \over 2}}


\output={\shipout\vbox{\makeheadline
                                      \ifnum\the\pageno>1 {\hrule}  \fi 
                                      {\pagebody}   
                                      \makefootline}
                   \advancepageno}

\headline{\noindent {\ifnum\the\pageno>1 
                                   {\tiny \ptitle\hfil page~\the\pageno}\fi}}
\footline{}
\newcount\zz  \zz=0  
\newcount\q   
\newcount\qq    \qq=0  

\def\pref #1#2#3#4#5{\frenchspacing \global \advance \q by 1     
    \edef#1{\the\q}
       {\ifnum \zz=1 { %
         \item{[\the\q]} 
         {#2} {\bf #3},{ #4.}{~#5}\medskip} \fi}}

\def\bref #1#2#3#4#5{\frenchspacing \global \advance \q by 1     
    \edef#1{\the\q}
    {\ifnum \zz=1 { %
       \item{[\the\q]} 
       {#2}, {\it #3} {(#4).}{~#5}\medskip} \fi}}

\def\gref #1#2{\frenchspacing \global \advance \q by 1  
    \edef#1{\the\q}
    {\ifnum \zz=1 { %
       \item{[\the\q]} 
       {#2}\medskip} \fi}}

 \def\sref #1{~[#1]}

\def\references#1{\zz=#1
   \parskip=2pt plus 1pt   
   {\ifnum \zz=1 {\noindent \bf References \medskip} \fi} \q=\qq

\gref{\hellb}{H. Hellmann, Acta Physicochim. URSS {\bf 1}, 913 (1935); {\bf 4}, 225 (1936);
 {\bf 4}, 324 (1936); J. Chem. Phys. {\bf 3}, 61 (1935).}
\gref{\hellm}{H. Hellmann and W. Kassatotchkin, Acta Physicochim. URSS {\bf 5}, 23 (1936);
 J.Chem. Phys.{\bf 4}, 324 (1936).}
\pref{\adam}{J. Adamowski, Phys. Rev. A }{31}{43 (1985)}{}
\pref{\var}{Y. P. Varshni and R. C. Shukla, Rev Mod. Physi.} {35}{130 (1963)}{}
\pref{\gry}{V. K. Gryaznov, Zh. Eksp. Teor. Fiz.} {78}{573 1980}
{[Sov. Phys. JETP {\bf 51},288 (1980)].}
\pref{\alek}{V. A. Alekseev, V. E. Fortov and I. T. Yakubov, USP. Fiz Nauk}{}
{ {\bf 139}, 193 1983[ Sov. Phys.-USP. {\bf 26}, 99 (1983)]}{}
\bref{\gom}{P. Gombas}{Die Statistische Theorie des Atoms und ihre Anwendungen}
{Springer, Berlin, 1949}{p304}
\gref{\callw}{J. Callaway, Phys. Rev. 112, 322 (1958); G. J. Iafrate, J. Chem. Phys. {\bf 45},
 1072 (1966); J. Callaway and P. S. Laghos, Phys. Rev. {\bf 187}, 192 (1969);
 G. McGinn, J. Chem. Phys. {\bf 53}, 3635 (1970).}
\pref{\bebn}{S. Bebnarek, J. Adamowski, and M. Saffczy{\'{n}}ski, Solid State  Commun.}{21}{1 (1977)}{}
\gref{\poll}{J. Pollmann and H. Buttner, Phys. Rev. B {\bf 16}, 4480 (1977); H. Buttner and J. Pollmann, Physica(Utrecht){\bf 117}/ {\bf 118} B, 278 (1983).}
\pref{\ada}{J. Adamowski, in proceedings of the XII confrence on Physics of semiconducting compounds, Jaszowiec, Poland, 1982, Solineum, Wroclaw}{}{(1983), p.139 and unpublished}{}
\pref{\dutt}{R. Dutt. U. Mukherji, and Y. P. Varshni, Phys. Rev.}{34}{777 (1986)}{}
\pref{\duart}{J. P. Duarte and R. L. Hall, Can. J. Phys.}{69}{1362 (1991)}{}
\pref{\kwato}{M.G. Kwato Njock, M. Nsangou, Z. Bona, S.G. Nana Engo, 
and B. Oumarou, Phys. Rev. A}{61}{042105 (2000)}{}
\bref{\reed}{M. Reed and B. Simon}{Methods of Modern Mathematical Physics IV:
 Analysis of Operators}{Academic, New York, 1978}
{The min-max principle for the discrete spectrum is discussed on p75}
\pref{\rhalla}{R. L. Hall, J. Math. Phys.} {24} {324 (1983)}{}
\pref{\rhallb}{R.L Hall, J. Math. Phys. }{25} {2708 (1984)}{}
\pref{\rhallc}{R.L. Hall, J. Math. Phys.} {34} {2779 (1993)}{}
 }

 \references{0}    

\tr 
\ppl{CUQM-85}
\ppl{math-ph/0107015}
\ppl{July 2001}
\vskip 1.0 true in
\centerline{\trbig Spectral bounds for the Hellmann potential}
\vskip 0.5true in
\baselineskip 12 true pt 
\centerline{\bf Richard L. Hall and Qutaibeh D. Katatbeh}\medskip
\centerline{\sl Department of Mathematics and Statistics,}
\centerline{\sl Concordia University,}
\centerline{\sl 1455 de Maisonneuve Boulevard West,}
\centerline{\sl Montr\'eal, Qu\'ebec, Canada H3G 1M8.}
\vskip 0.2 true in
\centerline{email:\sl~~rhall@mathstat.concordia.ca}
\bigskip\bigskip

\baselineskip = 18true pt  
\centerline{\bf Abstract}\medskip
The method of potential envelopes is used to analyse the bound state spectrum of the Schr\"odinger Hamiltonian $H=-\Delta+V(r), $ where the Hellmann potential is given by $V(r)=-A/r+Be^{-Cr}/r$, $A$ and $C$ are positive, and $B$ can be positive or negative. We established simple formulas yielding upper and lower bounds for all the energy eigenvalues.
  
\medskip\noindent PACS~~03.65.Ge,~03.65.Pm,~31.15.Bs,~02.30.Mv.

\np
  \title{1.~~Introduction}

The Hellmann potential $V(r)$ given by
$$ V(r)=-A/r+Be^{-Cr}/r\eqno{(1.1)}$$
 has many applications in atomic physics and condensed-matter physics \sref{\hellb-\ada}. The Hellmann potential, with B positive, was suggested originally by Hellmann\sref{\hellb-\hellm} and henceforth called the Hellmann potential if B is positive or negative.
The Hellmann potential was used as a model for alkali hydride molecules\sref{\var}. It was used also to represent the electron-ion \sref{\gry-\alek} and electron core interaction\sref{\gom-\callw}. It has also been shown that the main properties of the effective two-particle interaction for charged particles in polar crystals may be  described by this potential\sref{\bebn-\ada}.

  \title{2.~~The discrete spectrum : Scaling}

Many authors have studied the eigenvalues generated by the Hellmann potential and have tried to estimate them\sref{\hellb-\kwato}. For example Adamoski\sref{\adam} used  a variational framework to obtain accurate eigenvalues. Dutt, Mukherji and Varshni\sref{\dutt} and Kwato~Njock $\it{et~ al}$\sref{\kwato} applied the method of \hy{large}{N} expansion to approximate the bound states energies. In this paper we present simple upper- and lower-bound formulas obtained by the use of the comparison theorem and the envelope method \sref{\reed-\rhallc}.

We first show that discrete eigenvalues exist for the Hellmann potential for all values of $A > 0, B, {\rm and}~ C > 0.$  This result allows us to transcend the limit $B < A$ assumed to be necessary in an earlier attempt at this problem by geometrical methods\sref{\duart}. Suppose that $B \leq 0,$ then we immediately have that $-(A+|B|)/r < V(r) < -A/r.$ Since both upper and lower bounds are Hydrogenic potentials with discrete eigenvalues, the same follows for $V(r).$  Now we suppose that $B > 0.$ In this case the concern is that, for sufficiently large $B,$ the positive term might dominate the Coulomb term.  We see that this does not happen by the following argument.  The function $r e^{-Cr}$ has maximum value $1/(eC).$ Hence, for $B > 0,$ we have $Be^{-Cr}/r < (B/eC)/r^2,$ and we conclude that $-A/r < V(r) < -A/r + (B/eC)/r^2.$ But the `effective potential' for the Hydrogenic Atom  in a state of orbital angular momentum $\ell$ is given by 
$$V_{\rm eff}(r) = -A/r + \ell(\ell+1)/r^2.\eqno{(2.1)}$$
\nl Hence, again, we see that $V(r)$ is bounded above and below by Hydrogenic potentials whose corresponding Hamiltonians have discrete eigenvalues. This establishes our claim.\medskip 

If we denote the eigenvalues of $H=-\omega \Delta +A/r+Be^{-Cr}$ by ${\cal E}(\omega,A,B,C)$, and consider a scale change of the form $s=r/{\sigma}$, and choose the scale ${\sigma}={\omega}/A,$ then it is easy to show that, \nl
$$ {\cal E}(\omega,A,B,C)=C^2\omega {\cal E}(1,{A\over \omega C},{B\over \omega C},1).\eqno{(2.2)} $$
 
 \noindent Hence, the full problem is now reduced to the simpler two-parameter problem
$$ H=-\Delta - \alpha /r+\beta e^{-r}/r, \quad{\cal E}={\cal E}(\alpha,\beta), \alpha> 0. \eqno{(2.3)}$$

  \title{3.~~Energy bounds by the  Envelope Method}

 The Comparison Theorem of quantum mechanics tells us that an ordering between potentials implies a corresponding ordering of the eigenvalues. The `envelope method' is based on this result and provides us with  simple formulas for lower and upper bounds\sref{\rhalla-\rhallc}. We need a solvable model which we can use as an envelope basis. The natural basis to use in the present context is the hydrogenic potential
$$  h(r)=-1/r.  \eqno{(3.1)} $$


\nl ~~The spectrum generated by the potential $h(r)$ may be represented exactly
by the semi-classical expression 
$${\cal E}_{n\ell}(v) = \min_{s > 0}\{s + v\bar{h}_{n\ell}(s)\},\eqno{(3.2)}$$
where the `kinetic potential' $\bar{h}_{n\ell}(s)$ associated with the potential
$h(r) = -1/r$ is given, in this case, exactly by $\bar{h}_{n\ell}(s) = -s^{\half}/(n+\ell).$ 
If we now consider a potential, such as $V(r)$, which is a smooth transformation $V(r) = g(h(r))$
of $h(r),$ then it follows that a useful approximation for the corresponding kinetic potential 
$\bar{f}_{n\ell}(s)$ is given by
$$ \bar{f}_{n\ell}(s)\approx g(\bar{h}_{n\ell}(s)) \eqno{(3.3)}$$

\noindent If g is convex in (3.3), we get lower bounds ($\simeq = \ge  $) for all n and $\ell,$ and if g is concave we get upper bounds ($\simeq = \le $) for all n and $\ell$. 
 
For the Hellmann potential, if we use the potential $ h=-1/r$ as the envelope basis, then the sign of $g''$ depends only on the sign of B. An elementary calculation shows that
$$g''(h)=-BC^2e^{(C/h)}/h^3 = BC^2r^3e^{-Cr}.\eqno{(3.4)}  $$  
Hence, $g$ is convex if $B>0$ or concave if $B<0.$ Thus in this application of the method we obtain upper energy bounds for $B<0$ and lower bounds for $B>0.$ The following remarks explain briefly how these results are obtained.    

We suppose for definiteness that the transformation g(h) is smooth and convex $\it{i.e}$ $g''>0$, then each tangent to g is an affine transformation of h of the form

$$ {V^{(t)}}(r)=a(t)+b(t)h(r),\eqno{(3.5)}$$
where the variables  $a(t)$ and $b(t)$ are chosen  such that the graph of the potential $V(r)$ lies above the graph of the potential $h(r),$ but it is tangential to it
at a point, say $r=t.$ That is to say 
$$V(t)= a(t)+b(t)h(t)\eqno{(3.6)}$$
$${\rm and}\quad V'(t)=b(t)h'(t). \eqno{(3.7)}$$
This means that the `tangential potential', $V^{(t)}(r)$, and its derivative agree with 
V(r) at the point of contact, $r=t.$ 

Thus, by substituting (3.3) in (3.2), we find  
$${\cal E}_{n\ell}\approx \min_{r>0}\{{s+g(s^{1/2}/{{(n+\ell)}})} \},\eqno{(3.8)} $$
which yields an upper bound if $B<0$ and a lower bound if  $ B>0$. This can be further simplified by changing the minimazation variable $s$ to $r$ by the relation, 
$$g(\bar{h}_{nl}(s))=g(-s^{1/2}/(n+\ell))=V(r),\eqno{(3.9)}$$ which, in turn, implies $s=(n+\ell)^2/r^2.$  
 Hence we obtain finally the following semi-classical eigenvalue formula involving the potential $V(r)$ itself
$${\cal E}_{n\ell}\approx \min_{r>0}\{(n+\ell)^2/r^2+V(r)\}.\eqno{(3.10)} $$

\np

  \title{4.~~Results and conclusion}
 We now have a simple formula (3.10) for lower and upper bounds to the eigenvalues for the Hellmann potential.  In Fig.(1) we plot the ground-state eigenvalue bound (full line) as a function of $B$ for the case $A = 2,$ $C = 1,$ along with the corresponding point results of Adamowski\sref{\adam} as hexagons, and some accurate numerical values (dashed line). It is clear from this figure that the simple approximation formula gives an accurate estimate of the eigenvlues which is an upper bound if $B < 0,$ and a lower bound when $B > 0,$ as predicted by the theory.

If we fix $A,B,$ and $C$ and consider the Hamiltonian $H=-\Delta +vV(r),$ with eigenvalues ${\cal E}(v)$, then from (3.10) we immediatly obtain the following explicit parametric equations for the corresponding energy curve $\{v, {\cal E}(v)\},$ namely
 
$${\eqalign{ v&={2(n+\ell)^2\over {r^3V'(r)}}\cr
{\cal E}(v)&={(n+\ell)^2\over r^2}+{2(n+\ell)^{2}V(r)\over{r^3V'(r)}}.\cr}}\eqno{(4.1)}$$

\nl In Fig.(2) we exhibit the corresponding graphs of the function ${\cal E}(v)/v^2$ for $B=+1$ and $B=-1,$ again with $A = 2,$ and $C = 1,$ along with accurate numerical data shown as a dashed curve.  The main point of this work is to show that by elementary geometric reasoning one can obtain simple semi-classical approximations for the eigenvalues.  These results are complementary to purely numerical solutions and have the advantage that they are expressed analytically and allow one to explore the parameter space without having to attend to the arbitrary additional parameters and considerations which necessarily accompany numerical approaches with the aid of a computer.

   \title{Acknowledgments}
Partial financial support of this work under Grant No.GP3438 from the Natural Sciences and Engineering Research Council of Canada is gratefully acknowledged.\bigskip   
\np
\references{1}
\np
\hbox{\vbox{\psfig{figure=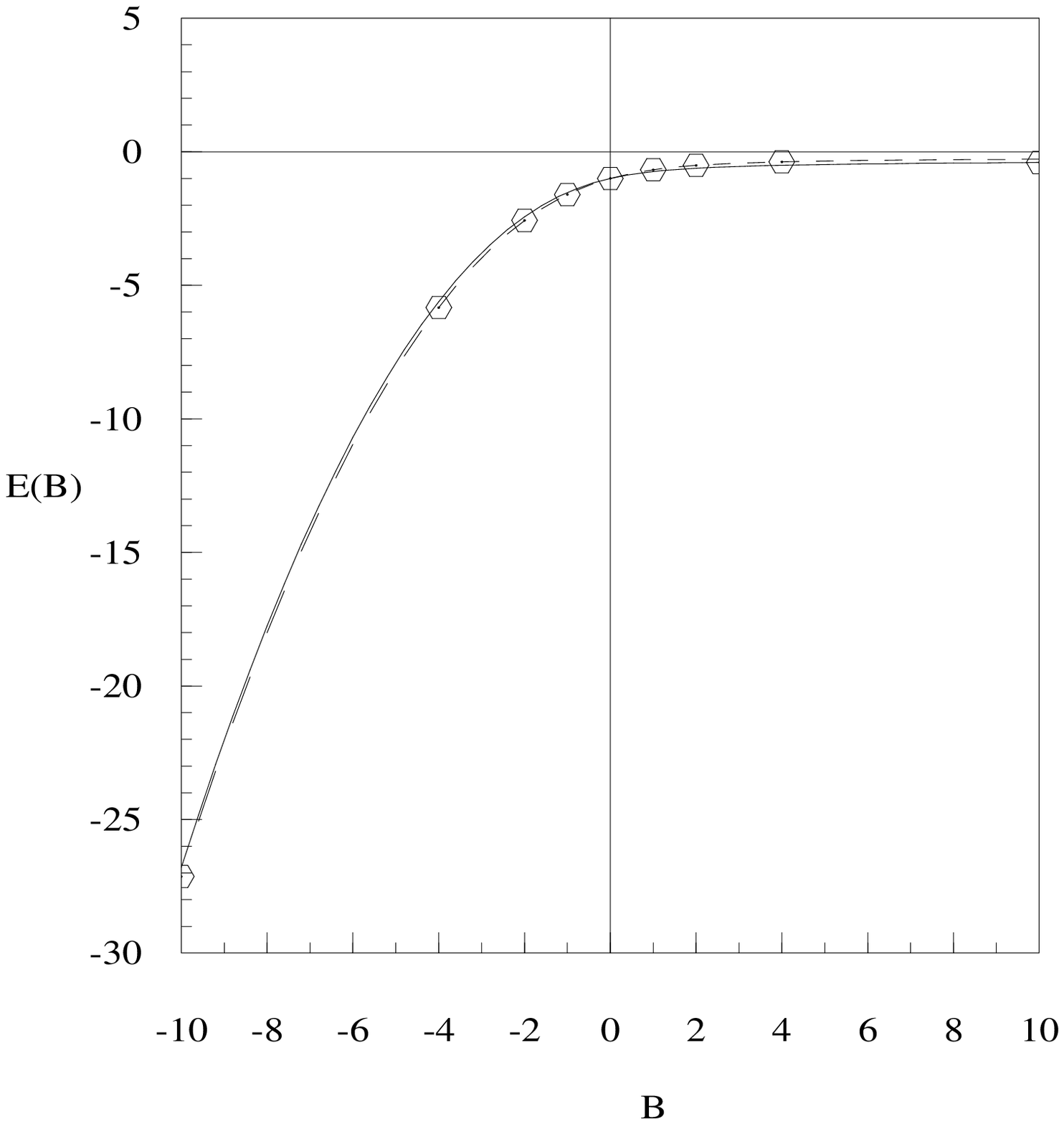,height=6in,width=5in,silent=}}}

\nl Fig.(1)~~The eigenvalues ${\cal E}(B)$ of the Hamiltonian $H=-\Delta-2/r+Be^{-r}/r$ for $n=1$ and $\ell=0.$ The continuous curve shows the bounds given by the formula (3.10), the dashed curve represents accurate numerical data, and the hexagons are the results of Adamowski\sref{\adam}. It is clear that the formula provides us with upper bounds when $B<0$ and lower bounds when $B>0$.\medskip

\np
\hbox{\vbox{\psfig{figure=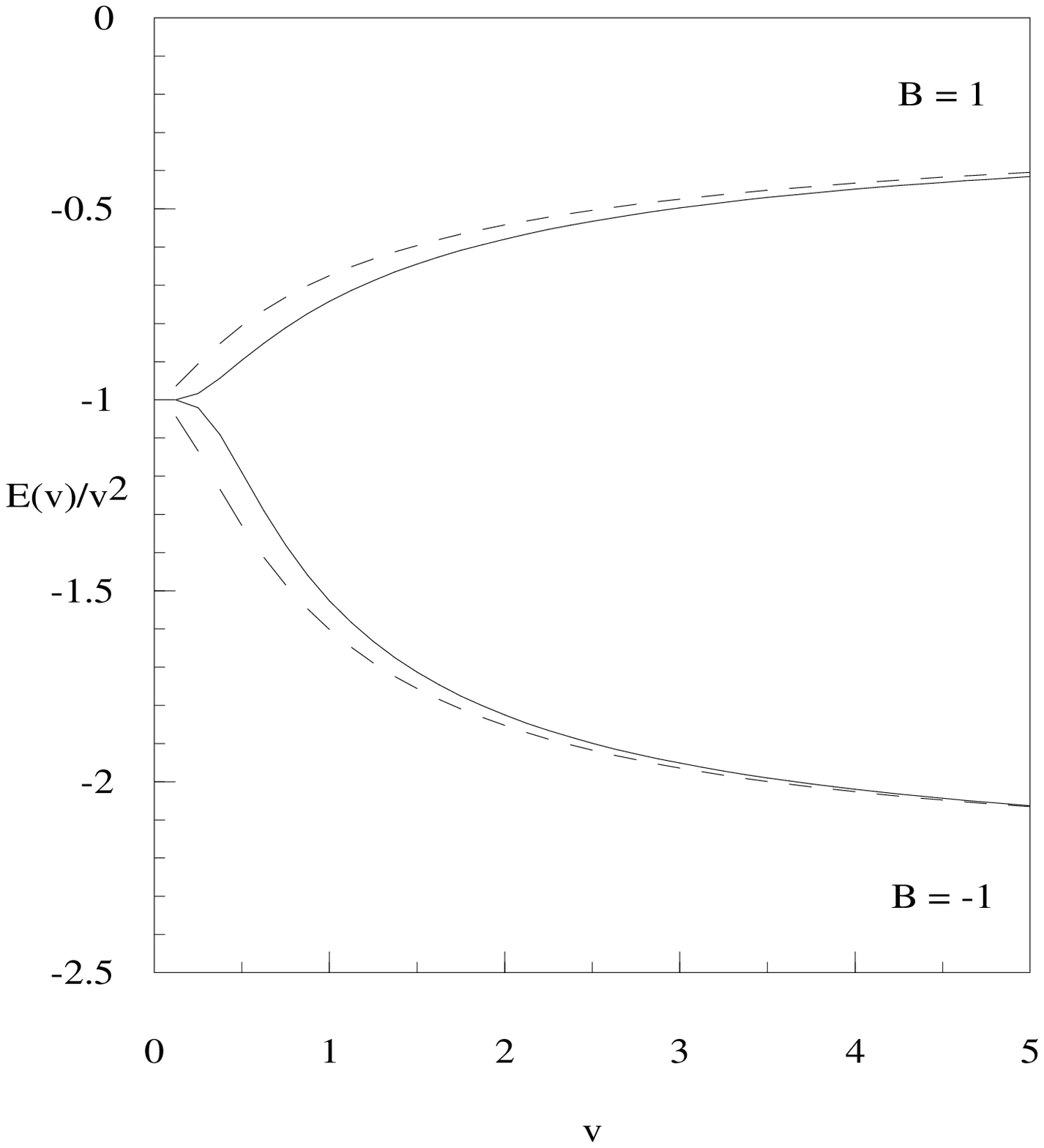,height=6in,width=5in,silent=}}}

\nl Fig.(2)~~The eigenvalue bounds (full-line) for ${\cal E}(v)/v^2$, where ${\cal E}(v)$ is the ground-state eigenvalue of the Hamiltonian $H=-\Delta+vV(r)$, for $A=2,$ $C=1,$ and $B=+1,-1,$ together with accurate numerical data (dashed-line). The parametric equations (4.1) yield upper bounds when $B<0,$ and lower bounds when $B>0.$

\hfil\vfil
\end